\documentclass[aip,bibnotes,numerical,reprint,floatfix,amssymb,amsmath]{revtex4-1}
\usepackage{graphicx}
\usepackage{bm}
\usepackage{txfonts}
\usepackage{hyperref}

\newcommand{\re}{\mathrm{e}}

\DeclareMathOperator{\Area}{Area}

\begin{document}

\title{Stretching and folding versus cutting and shuffling: An illustrated perspective on mixing and deformations of continua}

\author{Ivan C. Christov}
\email{christov@u.northwestern.edu}
\affiliation{Department of Engineering Sciences and Applied Mathematics, Northwestern University, Evanston, Illinois 60208, USA}

\author{Richard M. Lueptow}
\email{r-lueptow@northwestern.edu}
\affiliation{Department of Mechanical Engineering, Northwestern University, Evanston, Illinois 60208, USA}

\author{Julio M. Ottino}
\email{jm-ottino@northwestern.edu}
\thanks{To whom correspondence should be addressed.}
\affiliation{Department of Chemical and Biological Engineering, Northwestern Institute on Complex Systems, and Department of Mechanical Engineering, Northwestern University, Evanston, Illinois 60208, USA}


\begin{abstract}
We compare and contrast two types of deformations inspired by mixing applications -- one from the mixing of fluids (stretching and folding), the other from the mixing of granular matter (cutting and shuffling). The connection between mechanics and dynamical systems is discussed in the context of the kinematics of deformation, emphasizing the equivalence between stretches and Lyapunov exponents. The stretching and folding motion exemplified by the baker's map is shown to give rise to a dynamical system with a positive Lyapunov exponent, the hallmark of chaotic mixing. On the other hand, cutting and shuffling does not stretch. When an interval exchange transformation is used as the basis for cutting and shuffling, we establish that all of the map's Lyapunov exponents are zero. Mixing, as quantified by the interfacial area per unit volume, is shown to be exponential when there is stretching and folding, but linear when there is only cutting and shuffling. We also discuss how a simple computational approach can discern stretching in discrete data.
\end{abstract}

\maketitle

\section{Introduction}
\label{sec:intro}

The essence of mixing of a fluid with itself can be understood in terms of an array of striations of, say, two different colors of the same fluid (or two different fluids such as coffee and cream) undergoing stretching and folding. On top of stretching and folding we may superimpose diffusion, reaction, and, in special circumstances, breakup processes leading to droplet formation.\cite{omtfjk92,o89} This approach is the backbone of lamellar models of mixing. A fundamental measure of the quality of mixing is $a_V$, the interfacial area per unit volume of the striations (lamella or layers). Let $S$ be the interfacial area between fluid layers within a volume $V$ enclosing the point $\bm{x}$ at time $t$, then the interfacial area per unit volume is given by\cite{o89}
\begin{equation}
a_V(\bm{x},t) = \lim_{V\to 0} \frac{S}{V}.
\label{eq:av}
\end{equation}
A larger $a_V$ corresponds to better mixing.

We can imagine many iterative mixing protocols that generate large values of $a_V$ and create striations of the material of continually decreasing thickness in time. For fluids, a multitude of clever mixing designs can lead to the thinning of lamella, many inspired by a direct correspondence between the kinematics of mixing and chaotic dynamical systems.\cite{grbgh06} The simplest representation of mixing in terms of stretching and folding is the Smale horseshoe map, which stretches out a piece of material and folds it onto itself to form the shape of a horseshoe. A limiting case of this procedure is a map that stretches, cuts and re-stacks to generate interfacial area, the baker's transformation, named after the process by which a baker kneads dough.

Granular mixing has been studied as well, but less extensively than fluid mixing. In many respects, the ideas applied to fluids carry over to granular matter.\cite{mlo07} A key difference between the two is that granular flows may present surfaces of discontinuity, such as the interface between a flowing surface layer and the underlying static bed of granular material in an avalanche.\cite{r00} This new aspect of the flow leads to different models for the kinematics. In particular, mixing in granular flows in rotating containers\cite{Note1} (``tumblers'') can be thought of as ``cutting and shuffling,''\cite{smow08} a process different from stretching and folding.\cite{ank02} More on the fascinating behavior of granular matter can be found in Ref.~\onlinecite{k05}.

Stretching is a fundamental concept in mechanics and is covered in every continuum mechanics textbook in the context of kinematics (see, for example, the classic volumes of Truesdell\cite{t91} and Gurtin\cite{g81}), where it is identified with shear or extensional strain. Cutting and shuffling, in contrast, has been explored only recently. To illustrate the fundamental difference between the mixing mechanisms of stretching and folding versus cutting and shuffling, we compare two types of simple idealized mixing protocols: the well-known baker's map and  cutting and shuffling maps based on interval exchange transformations. Along the way we discuss the elegant connection between continuum mechanics and dynamical systems and show how some of the calculations of the relevant kinematic quantities, such as the deformation gradient and the principal stretches, are performed. Pertinent concepts from the theory of mixing are also reviewed within this context. We then discuss how to discern and measure stretching (or lack thereof) in practice.

\section{Kinematics of deformation}
\label{sec:kinematics}

We restrict our discussiuon to two spatial dimensions for simplicity.\cite{Note2} Consider a motion $\mathbf{\Phi}$ from the undeformed (reference or initial) configuration of the continuum (body) $\mathcal{B}_0$ to the deformed (current or final) configuration $\mathcal{B}_t$ at time $t$, both of which are regions in the Euclidean plane. Keeping with the mathematics notation, in this paper, we denote sets by capital calligraphic letters. Typically, we concern ourselves with motions that map the body back into itself, that is, $\mathcal{B}_t = \mathcal{B}_0$, but this restriction is not important for what follows. We are interested in discrete-time motions such as a repetitive mixing protocol. That is, we allow only $t=nT$, where $T$ is the duration (period) of the motion and $n$ is a positive integer. Then, we may write the motion as a map
\begin{equation}
\mbox{$\mathbf{\Phi}$:}\quad \mathcal{B}_0\to\mathcal{B}_t
\label{eq:motion1}
\end{equation}
such that after one iteration every $\bm{X}$ in $\mathcal{B}_0$ is mapped to an $\bm{x} = \mathbf{\Phi}(\bm{X})$ in $\mathcal{B}_t$, where by $\bm{x} \equiv (x,y)^\top$ we shall denote the position vector in the deformed configuration and $\bm{X} \equiv (X,Y)^\top$ represents the coordinates in the undeformed configuration. The $\top$ superscript denotes the transpose, meaning that $(x,y)$ is a {row} vector and $(x,y)^\top$ is a {column} vector. Unless otherwise noted, all vectors not written out in component form are considered to be column vectors.

We can now define the {deformation gradient} (or {Jacobian} matrix of the map) as
\begin{equation}
\mathbf{F} = \left(\nabla \bm{x}\right)^\top \equiv \dfrac{\partial(x,y)}{\partial(X,Y)} \equiv \begin{pmatrix} \dfrac{\partial x}{\partial X} & \dfrac{\partial x}{\partial Y} \\ \dfrac{\partial y}{\partial X} & \dfrac{\partial y}{\partial Y} \end{pmatrix}.
\label{eq:dg_matrix}
\end{equation}
It is typically assumed that the map is invertible and differentiable a sufficient number of times so that $\mathbf{F}$ exists and $0<\det\mathbf{F}<\infty$.\cite{t91,g81} In this paper we relax the differentiability assumption to consider a wider (and, arguably, more interesting) class of motion.

The {polar decomposition theorem}\cite{t91,g81,o89,hj85} allows us to write
\begin{equation}
\mathbf{F}=\mathbf{R}\mathbf{U},
\end{equation}
where $\mathbf{U}$ is a symmetric positive definite matrix\cite{Note3} (due to the assumption that $\det\mathbf{F}>0$) and $\mathbf{R}$ is a proper-orthogonal matrix.\cite{Note4} That is, we can locally decompose the deformation into a {rotation} $\mathbf{R}$ and a {stretch} $\mathbf{U}$. Because the matrix $\mathbf{U}$ is symmetric positive definite, it has an orthonormal basis of eigenvectors $\{\bm{e}_1,\bm{e}_2\}$ and strictly positive real eigenvalues $\{\sigma_1,\sigma_2\}$ satisfying
\begin{equation}
\mathbf{U}\bm{e}_i = \sigma_i\bm{e}_i.
\end{equation}
The eigenvalues $\sigma_i$ and eigenvectors $\bm{e}_i$ are called, respectively, the {principal stretches} and the {principal directions} because an infinitesimal line segment of length $d\ell$ oriented in the direction $\bm{e}_i$ has length $\sigma_i d\ell$ after undergoing the motion (that is, after the map $\mathbf{\Phi}$ is applied). Hence, $\sigma_i > 1$ for stretching and $0< \sigma_i < 1$ for compression along the direction $\bm{e}_i$.

{\it Problem 1}. Consider $d\bm{x} = \mathbf{\Phi}(\bm{X} + d\bm{X}) - \mathbf{\Phi}(\bm{X})$, and show that $d\bm{x} = \mathbf{F}d\bm{X}+$ higher order terms when $\|d\bm{X}\| \ll 1$. (Hint: use Taylor's theorem for a vector function.) Here, $\|d\bm{X}\|\equiv\sqrt{d\bm{X} \cdot d\bm{X}}$ is the norm induced by the usual Euclidean dot (inner) product. Then let $d\bm{X} = (d\ell)\bm{e}_i$ and show that $\|d\bm{X}\| = d\ell$ and $\|d\bm{x}\| = \sigma_i d\ell$.

The stretches may depend on the coordinate $\bm{X}$ in the undeformed configuration and also on time (or, in the present context, the number of times $n$ that the mixing protocol is repeated). Typically, we are concerned with the {largest} principal stretch $\hat{\sigma} = \max\{\sigma_1,\sigma_2\}$, which is a scalar field $\hat{\sigma}=\hat{\sigma}(\bm{X};n)$ that describes the stretching experienced by the body due to its motion. In practice, it is convenient to calculate the eigenvalues $\{\kappa_1,\kappa_2\}$ of the (right) {Cauchy--Green strain tensor} $\mathbf{C} \equiv \mathbf{F}^\top\mathbf{F}$ instead. These are just the squares of the principal stretches. In this way $\hat{\sigma}$ can be computed without explicitly finding the polar decomposition of $\mathbf{F}$.

{\it Problem 2}. Show that $\kappa_i = \sigma_i^2$ using the definition of $\mathbf{C}$, the polar decomposition theorem, and the properties of eigenvalues.

\section{Dynamical systems framework of kinematics}

Equation~\eqref{eq:motion1} also defines a {dynamical system}, which in the most general sense is defined as a rule of evolution on a state space (the body).\cite{m07} This connection between the kinematics of continua and dynamical systems has been successfully exploited in the study of both fluid mixing\cite{o89} and granular mixing.\cite{mlo07}

There is a direct correspondence between the languages of dynamical systems and continuum mechanics, the most important of which, for the present purposes, is the correspondence between stretches and \emph{Lyapunov (characteristic) exponents}. For a discrete-time map, these exponents are defined (see, for example, Ref.~\onlinecite{sow06}, Sec.~5.3.1) as
\begin{equation}
\lambda(\bm{X},\bm{v}) = \lim_{n\to\infty} \frac{1}{n} \ln \| (\nabla \mathbf{\Phi}^n )^\top \bm{v} \|,
\label{eq:lyap_exp}
\end{equation}
where $\mathbf{\Phi}^n \equiv \mathbf{\Phi}\circ\cdots\circ\mathbf{\Phi}$ ($n$ compositions of the map). The Lyapunov exponents depend on the position $\bm{X}$ and on the direction $\bm{v}$ ($\|\bm{v}\| =1$). The quantity $\nabla \mathbf{\Phi}^n$ can be calculated by the chain rule along a trajectory starting at a given $\bm{X}$ in $\mathcal{B}_0$. If $\mathbf{F}\equiv(\nabla \mathbf{\Phi})^\top$ happens to be independent of $\bm{X}$, we have $(\nabla \mathbf{\Phi}^n)^\top = \mathbf{F}^n$. Lyapunov exponents are important in the context of the asymptotic stability of infinitesimal perturbations and, provided certain conditions are satisfied, can be interpreted as the growth (or decay) rates of these perturbations {along a trajectory}.

A related concept is the \emph{finite-time} Lyapunov exponents defined as
\begin{equation}
\gamma(\bm{X},\bm{v};n) = \frac{1}{n} \ln \| (\nabla \mathbf{\Phi}^n )^\top \bm{v} \|.
\label{eq:ftle}
\end{equation}
Note that if $(\nabla \mathbf{\Phi}^n)^\top = \mathbf{F}^n$, we have
\begin{equation}
\max_{\bm{v}\ne\bm{0},\,\,\|\bm{v}\|=1} \gamma(\bm{X},\bm{v};n) = \frac{1}{n} \ln \sqrt{\rho\Big( (\mathbf{F}^n )^\top\mathbf{F}^n \Big)},
\label{eq:stretch_rr}
\end{equation}
where $\rho(\mathbf{A})$ denotes the \emph{spectral radius} of the matrix $\mathbf{A}$, that is, its largest eigenvalue in absolute value. As we discussed at the end of Sec.~\ref{sec:kinematics}, the square root of the largest eigenvalue of $(\mathbf{F}^n )^\top\mathbf{F}^n$ is the largest principal stretch (for the deformation resulting from applying the map $n$ times). Therefore, there is a one-to-one relation between the largest finite-time Lyapunov exponent and the largest principal stretch:
\begin{equation}
\gamma^{\max}(\bm{X};n) \equiv \max_{\bm{v}\ne\bm{0},\,\,\|\bm{v}\|=1} \gamma(\bm{X},\bm{v};n) = \frac{1}{n}\ln\hat{\sigma}(\bm{X};n).
\label{eq:lyap_stretch}
\end{equation}
If the limit in Eq.~\eqref{eq:lyap_exp} exists, then this relation carries over to the largest (infinite-time) Lyapunov exponent as well.\cite{Note5}

{\it Problem 3}. Derive Eq.~\eqref{eq:stretch_rr} using the Rayleigh--Ritz theorem from linear algebra.\cite{hj85} Hint: write out the norm in the definition of $\gamma$ as a dot product and re-arrange terms. Also, you are allowed to bring the max inside any logarithm or square root because these functions are monotonically increasing.

We will call a dynamical system with a positive Lyapunov exponent \emph{chaotic} (although there are variety of sometimes equivalent ways to define chaos\cite{bc96}). If a motion in the sense of continuum mechanics stretches the continuum ($\hat{\sigma} > 1$), the corresponding dynamical system has a positive Lyapunov exponent by virtue of Eq.~\eqref{eq:lyap_stretch}. Note that, in this context, the chaotic dynamics occur in {physical space}, not in {phase space} as is the case for the dynamical systems of classical mechanics.\cite{smf90}

There is a temptation to equate stretching and folding motions (equivalently, ones that give rise to chaotic dynamical systems) with efficient generation of $a_V$. However, as we will show in the following examples, there are other possibilities for generating $a_V$ without stretching and folding.

\section{Example mixing protocols}
\label{sec:maps}
One of the best known stretching and folding operations is the baker's map,\cite{o89,sow06} which can be found as early as 1951 in the mixing literature\cite{sw51} and 1937 in the mathematics literature.\cite{h37} In contrast, cutting and shuffling, two examples of which we construct using interval exchange transformations, is a type of mixing protocol that has been proposed only recently.\cite{smow08,col10,col10b,jlosw10} Suppose we begin with a square region, half of which is filled with a while material while the other half is filled with a gray material (see Fig.~\ref{fig:maps}). Then, from an inspection of the initial (undeformed) and final (deformed) configurations of the continuum undergoing deformation described by these protocols, it is clear that two of them---the baker's map (BM) and the second cutting and shuffling map (CS2)---lead to mixing of the gray and white materials. (The map CS1 does not produce further striations of material, as it only re-arranges the layers produced by the first set of cuts in a periodic manner.) However, the following simple analytic calculations, which are confirmed by numerical calculations in Sec.~\ref{sec:numerics}, show that only the baker's map stretches the material continuum and gives rise to a chaotic dynamical system in the usual sense.

\subsection{The baker's map}
\label{sec:bakers}
\begin{figure*}[!ht]
\includegraphics[width=\textwidth]{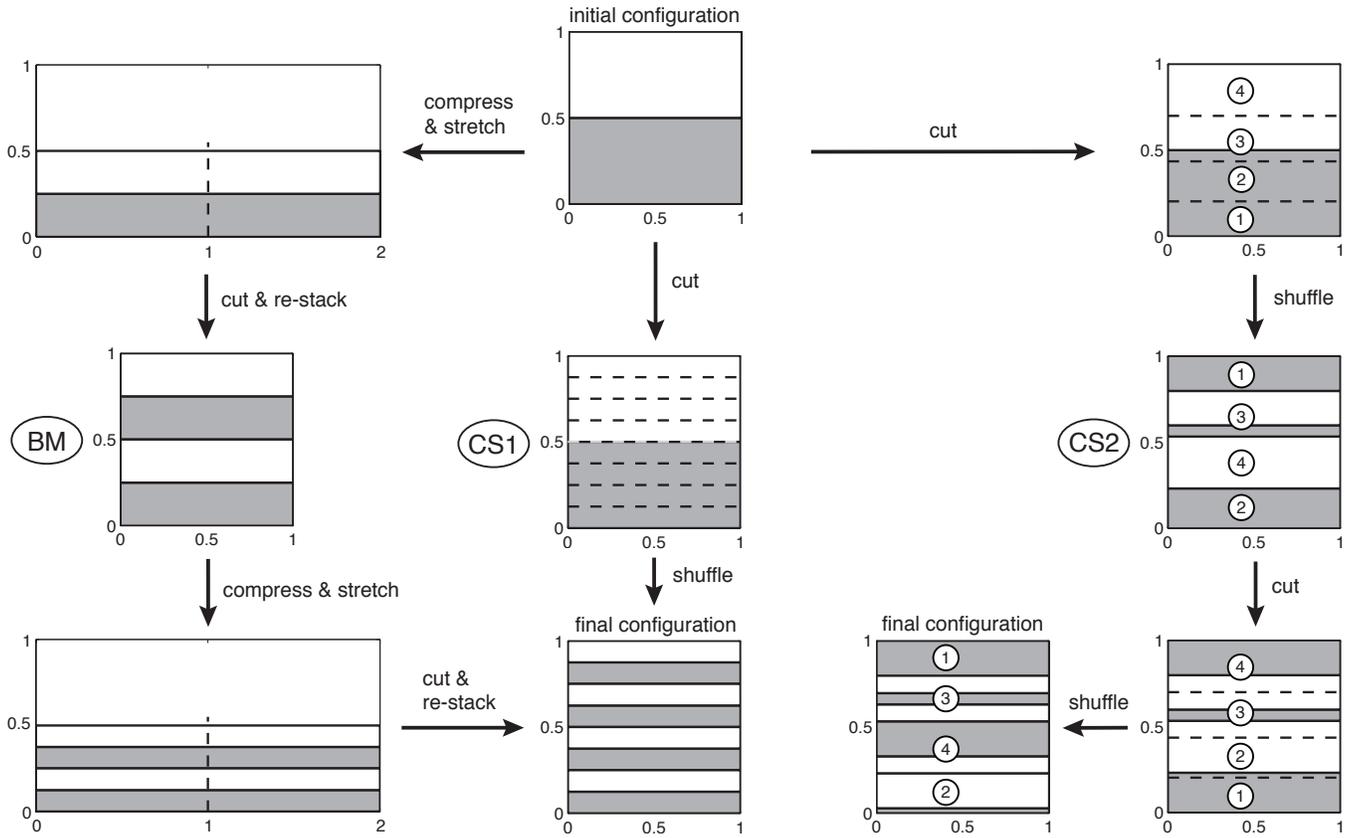}
\caption{Illustration of mixing by three different protocols. The left path (BM) shows two applications of the baker's map, which stretches the underlying continuum. The middle path (CS1) is one iteration of a cutting and shuffling map with $k=8$ intervals and length ratio $r=1$, which does not stretch and leads to the same macroscopic picture as the left path (BM). The right path (CS2) is two applications of a cutting shuffling map based on an interval exchange transformation with $k=4$ and $r=\sqrt{1.3}$, which also does not stretch but results in a different macroscopic picture. Only the maps BM and CS2 produce an increasing number of finer striations of each color upon repeated application.}
\label{fig:maps}
\end{figure*}

We let $\mathcal{B}_t = \mathcal{B}_0 = [0,1]^2$, and write the baker's map $\mathbf{\Phi}_{\mathrm{BM}}$: $[0,1]^2\to[0,1]^2$ as
\begin{equation}
\mathbf{\Phi}_{\mathrm{BM}}(\bm{X}) =
\begin{cases}
\left(2X,\dfrac{1}{2}Y\right)^\top, & \left (0\le X< \dfrac{1}{2} \right);\\
\left(2X-1,\dfrac{1}{2}Y+\dfrac{1}{2}\right)^\top, & \left(\dfrac{1}{2} \le X \le 1 \right).
\end{cases}
\label{eq:baker}
\end{equation}
Simply put, this map involves compressing the unit square to half its height, stretching it to twice its width, cutting vertically the resulting rectangle in half along $X=1/2$ and stacking the pieces [see Fig.~\ref{fig:maps} (BM)] -- much like how a baker kneads dough or a taffy pull machine makes candy.\cite{Note6} Other applications include the Kenics mixer\cite{gapm03} and the related partitioned pipe mixer (see, for example, Ref.~\onlinecite{o89}, Sec.~8.2). The baker's map has even been used to explain the movement of bubbles in a foam network.\cite{tt08} In fluid mixing, the cut and re-stack step cannot be accomplished exactly, but cutting and re-stacking is precisely what happens in the extrusion of multi-layer plastics.\cite{cm00}

Even though the baker's map constitutes stretching, cutting and re-stacking, it possesses (as we will show) the essential property of a stretching and folding motion -- stretching in the continuum mechanics sense ($\hat{\sigma}>1$) or, equivalently, a positive Lyapunov exponent. The Smale horseshoe map\cite{o89,d03,wo04} is classical example of stretching and folding. However, the horseshoe map does not preserve area, which has significant implications for the types of chaos exhibited by the dynamical system because it allows for the existence of \emph{attractors}.\cite{ms92} We avoid such complications by idealizing the Smale horseshoe map as the baker's map.

From Eqs.~\eqref{eq:dg_matrix} and \eqref{eq:baker} it follows that
\begin{equation}
\mathbf{F}_{\mathrm{BM}} = \begin{pmatrix} 2 & 0 \\ 0 & \dfrac{1}{2} \end{pmatrix} \qquad \left(X\ne\dfrac{1}{2}\right),
\label{eq:baker_dg}
\end{equation}
and $\mathbf{F}_{\mathrm{BM}}$ is undefined at the cut $X=1/2$. The polar decomposition of this matrix is simple: $\mathbf{R}_{\mathrm{BM}} = \mathbf{I}$ and $\mathbf{U}_{\mathrm{BM}} = \mathbf{F}_{\mathrm{BM}}$. The eigenvalues of $\mathbf{U}_{\mathrm{BM}}$ are $\sigma_1=2$ and $\sigma_2=1/2$. Therefore, the largest principal stretch is $\hat{\sigma} = 2 > 1$ everywhere after one iteration of the protocol ($n=1$). Clearly, the baker's map stretches the underlying material continuum.

Because $\mathbf{F}_{\mathrm{BM}}$ is independent of $\bm{X}$, iterating the baker's map results in
\begin{equation}
\left(\nabla\mathbf{\Phi}_{\mathrm{BM}}^n\right)^\top = \mathbf{F}_{\mathrm{BM}}^n = \begin{pmatrix} 2 & 0 \\ 0 & \dfrac{1}{2} \end{pmatrix}^n = \begin{pmatrix} 2^n & 0 \\ 0 & \dfrac{1}{2^n} \end{pmatrix} \qquad \big(\bm{X}\notin\mathcal{C}_{\mathrm{BM}}^n\big),
\label{eq:baker_dgn}
\end{equation}
and $\mathbf{F}_{\mathrm{BM}}$ is undefined along the cuts, that is, the set of points
\begin{equation}
\mathcal{C}_{\mathrm{BM}}^n = \{\bm{X}\;\text{in}\;[0,1]^2\,|\, X = i/2^n,\; i=1,2,\hdots,2^n-1;\; 0\le Y\le 1\}.
\label{eq:baker_cuts}
\end{equation}
Once again, $\mathbf{R}_{\mathrm{BM}}^n = \mathbf{I}$ and $\mathbf{U}_{\mathrm{BM}}^n = \mathbf{F}_{\mathrm{BM}}^n$. Consequently, the largest principal stretch of the deformation is
\begin{equation}
\hat{\sigma}_{\mathrm{BM}}(\bm{X};n) = 2^n \qquad(\bm{X}\notin\mathcal{C}_{\mathrm{BM}}^n).
\label{eq:baker_stretch}
\end{equation}

Turning to the Lyapunov exponents, the two directions of interest are the principal directions $\bm{e}_i$ (the eigenvectors of $\mathbf{U}_{\mathrm{BM}}$), which are $\bm{e}_1 = (1,0)^\top$ and $\bm{e}_2 = (0,1)^\top$. Therefore, upon multiplying the vectors $\bm{e}_1$, $\bm{e}_2$ by the matrix in Eq.~\eqref{eq:baker_dgn} and taking the norm of the resulting vector, $\ln\|\mathbf{F}_{\mathrm{BM}}^n\bm{e}_{1,2}\| = \pm n\ln 2.$ We substitute this expression into Eq.~\eqref{eq:ftle} and obtain the finite-time Lyapunov exponents (along the principal directions) of the baker's map
\begin{equation}
\gamma_{\mathrm{BM}}(\bm{X},\bm{e}_{1,2};n) = \pm\ln 2 \qquad(\bm{X}\notin\mathcal{C}_{\mathrm{BM}}^n),
\label{eq:ftle_baker}
\end{equation}
and they do not exist for $\bm{X}$ in $\mathcal{C}_{\mathrm{BM}}^n$.

{\it Problem 4}. Show that the larger finite-time Lyapunov exponent in Eq.~\eqref{eq:ftle_baker} can also be calculated using Eqs.~\eqref{eq:lyap_stretch} and \eqref{eq:baker_stretch}.

The Lyapunov exponents defined in Eq.~\eqref{eq:lyap_exp} can be obtained by taking the limit $n\to\infty$ in Eq.~\eqref{eq:ftle_baker}:
\begin{equation}
\lambda_{\mathrm{BM}}(\bm{X},\bm{e}_{1,2}) = \pm\ln 2 \qquad \big(\bm{X}\notin\mathcal{C}_{\mathrm{BM}}^\infty\big).
\label{eq:le_baker}
\end{equation}
A subtle but important point\cite{Note7} is that the set of cuts $\mathcal{C}_{\mathrm{BM}}^n$ is a null subset of the domain $\mathcal{B}_0$ as $n\to\infty$\cite{wo04} because it is a {set of measure zero}. Thus the quantities in Eqs.~\eqref{eq:ftle_baker} and \eqref{eq:le_baker} exist {almost everywhere}.

Since $\lambda_{\mathrm{BM}}(\bm{X},\bm{e}_{1})>0$, it follows that the baker's map gives rise to a chaotic dynamical system; that is, if we were to track an infinitesimal ball of material points of the continuum, they will spread exponentially fast from each other with repeated applications of the map because it possesses a positive Lyapunov exponent. The simplicity of the baker's map and its ability to render many analytical calculations tractable have made it one of the classic examples of a chaotic dynamical system.\cite{lb08} We refer the reader to, for example, Ref.~\onlinecite{tg06}, Sec.~5.1 for a thorough overview, including the non-area-preserving (dissipative) version of the map.

\subsection{Cutting and shuffling maps}
\label{sec:iet}
Cutting and shuffling is an operation well-known to card players and mathematicians. One of the simplest dynamical systems that can be successfully studied analytically is the interval exchange transformation; see, for example, Ref.~\onlinecite{kh95}, Sec.~14.5. An interval exchange transformation, which we write as $E_{S,\Pi}$: $\mathcal{I}\to\mathcal{I}$, replaces a deck of cards with a continuous interval $\mathcal{I}$ of the real line (we take $\mathcal{I}=[0,1]$ without loss of generality) and subdivides it into a collection of $k$ disjoint subintervals $S = \{\mathcal{I}_1,\hdots,\mathcal{I}_k\}$. These subintervals are translated (``shuffled'') according to a rule given by some permutation $\Pi$ of the integers between $1$ and $k$. Finally, the interval $\mathcal{I}$ is put back together as $\mathcal{I}=\mathcal{I}_{\Pi(1)}\cup\cdots\cup \mathcal{I}_{\Pi(k)}$. The unusual type of continuum motions encountered in tumbled granular flows\cite{smow08,col10,col10b,jlosw10} that we noted in Sec.~I are specific two-dimensional generalizations, called \emph{piecewise isometries},\cite{g02,d06} of these simple maps.

To mix the domain $\mathcal{B}_0=[0,1]^2$, we can apply an interval exchange transformation in the $Y$-direction and extend it in the $X$-direction by making each subinterval $\mathcal{I}_i$ into a rectangle of unit horizontal length. This construction is a special case of the more general class of \emph{rectangle exchange transformations}.\cite{h81} Consider the simple special case of $k=8$ equal subintervals shuffled according to the permutation $\Pi([12345678])=[15263748]$, numbering the subintervals from bottom to top so that intervals 1 through 4 are gray and intervals 5 through 8 are white. This interval exchange transformation is equivalent to cutting a deck of 8 cards exactly in half and shuffling them perfectly so that the bottom card from the first half (gray) is just below the bottom hard from the second half (white), and so on. As shown in the middle path of Fig.~\ref{fig:maps} (CS1), one iteration of this map results in layers of gray and white that appear identical to those produced by two iterations of the baker's map. It is not difficult to see how one iteration of the same cutting and shuffling map with, say, $k=64$ subintervals results in the same number and placement of layers of gray and white as $n=5$ iterations of the baker's map. However, repeated application of the map CS1 does not result in mixing for any choice of $k$. At $n=2$ (for $k=8$), demixing\cite{ol08} occurs: the number of distinct stripes decreases from 8 to 4. Fortunately, there is a well-developed theory to guide us in constructing interval exchange transformations that mix well.

To this end, consider the more general case in which the intervals are not equal and are not shuffled in such an orderly fashion. It is well known that for an interval exchange transformation to exhibit interesting behavior (specifically, to have no periodic orbits and to be {ergodic}\cite{Note8}), it must satisfy the \emph{Keane condition}.\cite{v06} This condition requires that the ratio of the lengths of adjacent intervals $|\mathcal{I}_{i}|/|\mathcal{I}_{i-1}|$ be an irrational number and that the permutation $\Pi$ be {irreducible}; that is, applying $\Pi$ to any of the subsets $\{1\}$, $\{1,2\}$, $\{1,2,3\}$ up to $\{1,2,\hdots,k-1\}$ does not yield a permutation of just the elements of the subset. For example, the permutation $\Pi([12345]) = [31254]$ is reducible (that is, not irreducible) because the first three elements are a permutation of only themselves (neither 1, 2, nor 3 maps to 4 or 5). In contrast, $\Pi([12345]) = [31524]$ is irreducible because $3$ maps to $5$. To satisfy the Keane condition, we consider $k=4$ subintervals, and suppose their lengths are chosen so that the first is $|\mathcal{I}_1|=\eta$ and each consecutive subinterval has length $r$ times the length of the previous one: $|\mathcal{I}_{i}|=r|\mathcal{I}_{i-1}|$. Because the length of $\mathcal{I}=[0,1]$ must be preserved, $\eta + r\eta + r^2\eta + r^3\eta = 1$. Given $r$ we can solve for $\eta$ from this relation. We take $r = \sqrt{1.3}$ and, numbering from bottom to top, $\Pi([1234]) = [2431]$ to satisfy the Keane condition. In terms of the continuum mechanics language introduced in Sec.~\ref{sec:kinematics}, the map takes the form
\begin{multline}
\mathbf{\Phi}_{\mathrm{CS2}}(\bm{X}) =\\
\begin{cases} \big(X, Y + (r+r^2+r^3)\eta\big)^\top, & (0\le Y < \eta)\\
\big(X, Y -\eta\big)^\top, & (\eta\le Y < [1+r]\eta)\\ \big(X, Y + (r^3-1)\eta\big)^\top, & ([1+r]\eta\le Y < [1+r+r^2]\eta)\\ \big(X,Y - (1+r^2)\eta\big)^\top, & ([1+r+r^2]\eta\le Y \le 1).\end{cases}
\label{eq:cs2_map}
\end{multline}
The action of this cutting and shuffling map, constructed from an interval exchange transformation, is illustrated by the right path in Fig.~\ref{fig:maps} (CS2).

{\it Problem 5}. Derive Eq.~\eqref{eq:cs2_map} by considering where the rectangles go in Fig.~\ref{fig:maps} (CS2). (Hint: rectangle 1 moves up by the height of rectangles 2, 3, and 4, that is, $(r+r^2+r^3)\eta$, etc.)

Because we are simply translating the strips in the vertical direction, calculating the deformation gradient from Eq.~\eqref{eq:dg_matrix} is trivial:
\begin{equation}
\mathbf{F}_{\mathrm{CS2}} = \begin{pmatrix} 1 & 0 \\ 0 & 1\end{pmatrix} \qquad \big(\bm{X}\notin\mathcal{C}^1_{\mathrm{CS2}}\big),
\label{eq:iet_dgn}
\end{equation}
and it is undefined along the cuts, that is, the set of points
$\mathcal{C}^1_{\mathrm{CS2}} =\{\bm{X}\;\text{in}\;[0,1]^2\,|\,0\le X\le1;\;Y = r^i\eta,\; i=0,1,2\}$.
Unlike for the baker's map, finding the set of cuts $\mathcal{C}^n_{\mathrm{CS2}}$ for any $n$ for the above cutting and shuffling map (CS2) is not trivial and, in general, an open problem.\cite{n09} Still, it is a countable set, and therefore $\mathcal{C}^\infty_{\mathrm{CS2}}$ is a null subset of $\mathcal{B}_0$.

{\it Problem 6}. Write a computer program that finds the number and location of the distinct cuts after $n$ applications of $\mathbf{\Phi}_{\mathrm{CS2}}$. What happens if you change $r$ or $\Pi$?

For any given trajectory that does not start at some $\bm{X}$ in $\mathcal{C}^n_{\mathrm{CS2}}$, it is clear that the deformation gradient after $n$ iterations is the product of identity matrices like the one in Eq.~\eqref{eq:iet_dgn}. Hence $\mathbf{F}_{\mathrm{CS2}}^n = \mathbf{I}$. Therefore, the principal stretches are $\sigma_{1} = \sigma_{2} = 1$, and
\begin{equation}
\hat{\sigma}_{\mathrm{CS2}}(\bm{X};n) = 1\quad(\bm{X}\notin\mathcal{C}_{\mathrm{CS2}}^n).
\label{eq:iet_stretch}
\end{equation}

From Eq.~\eqref{eq:ftle} and $\mathbf{F}_{\mathrm{CS2}}^n = \mathbf{I}$, we can calculate the finite-time Lyapunov exponents along the principal directions:
\begin{equation}
\gamma_{\mathrm{CS2}}(\bm{X},\bm{e}_{1};n) = \gamma_{\mathrm{CS2}}(\bm{X},\bm{e}_{2};n) = 0\qquad(\bm{X}\notin\mathcal{C}_{\mathrm{CS2}}^n),
\label{eq:ftle_iet}
\end{equation}
and they do not exist along the cuts. If we take the limit $n\to\infty$ in Eq.~\eqref{eq:ftle_iet}, we arrive at the Lyapunov exponents [Eq.~\eqref{eq:lyap_exp}]:
\begin{equation}
\lambda_{\mathrm{CS2}}(\bm{X},\bm{e}_{1}) = \lambda_{\mathrm{CS2}}(\bm{X},\bm{e}_{2}) = 0 \quad \big(\bm{X}\notin\mathcal{C}^\infty_{\mathrm{CS2}}\big).
\label{eq:ftle_iet2}
\end{equation}
Equations~\eqref{eq:ftle_iet} and \eqref{eq:ftle_iet2} are in stark contrast with Eqs.~\eqref{eq:ftle_baker} and \eqref{eq:le_baker}. In addition, this type of argument can be formalized into a proof showing that all Lyapunov exponents are zero for the more general class of piecewise isometry maps on the plane.\cite{fd08}

{\it Problem 7}. Show that the finite-time Lyapunov exponents in Eq.~\eqref{eq:ftle_iet} can also be obtained from Eqs.~\eqref{eq:lyap_stretch} and \eqref{eq:iet_stretch}.

Therefore, the map $\mathbf{\Phi}_{\mathrm{CS2}}$ (and, by the same logic, the map $\mathbf{\Phi}_{\mathrm{CS1}}$) does not give rise to chaotic dynamics. Specifically, an infinitesimal ball of material points in the continuum does {not} spread apart under $n$ iterations of this map (for any choice of $n$), unless the ball happens to overlay a cut. Hence, the distance between any two points in $\mathcal{B}_0$ can grow only if they become separated by a cut.

We have assumed that the {same} specific shuffle (defined by the given permutation $\Pi$) is performed {once} at each iteration of the map. In doing so, we did not consider the possibility that a white piece of material may be moved next to another white one, meaning that the shuffle is not necessarily optimal in terms of mixing. Thus, an improvement would be to allow multiple shuffles to be performed at each step of the protocol. There is an elegant theory\cite{ad86,tt00} (in the context of shuffling decks of cards) that gives an estimate of how many such shuffles are required for the stack to be ``sufficiently random.''

\section{Defining and quantifying mixing}
\label{sec:mixing}

The goal of the protocols we have considered is to mix the gray with the white in Fig.~\ref{fig:maps}, but what do we mean by ``mix''? If we go by the earlier intuitive definition, that is, to mix is to generate interfacial area $a_V$, then one thing is clear about the baker's map protocol from Sec.~\ref{sec:bakers}: the number of stripes (of unit length) in the final configuration is $2^{n+1}$. Thus, the bulk interfacial area between the gray and white materials is
\begin{equation}
a_V = 2^{n+1} - 1 = 2\re^{hn} - 1.
\label{eq:bakers_av}
\end{equation}
By expressing $a_V$ as an exponential function of $n$ with growth rate $h = \ln 2 = \gamma^{\max}_{\mathrm{BM}}$,\cite{Note9} we are able to illustrate the well-known result that chaotic advection, through stretching and folding, generates $a_V$ at an exponential rate.

For the cutting and shuffling maps from Sec.~\ref{sec:iet}, the most interfacial area that can be generated in one iteration is equal to the number of cuts, which is $k-1$ for $k$ subintervals. Hence, assuming that the interface at $Y=1/2$ in the initial configuration does not coincide with a cut, the upper bound is
\begin{equation}
a_V \le 1 + (k-1)n.
\label{eq:iet_av}
\end{equation}
Equation~\eqref{eq:iet_av} is only an upper bound because not every cut gives rise to an intermaterial interface after shuffling. For example, while on the first iteration of the map CS1 in Fig.~\ref{fig:maps} the seven cuts produce seven interfaces, on the second iteration only three interfaces remain though we have made seven fresh cuts. The nature of the upper bound in Eq.~\eqref{eq:iet_av} illustrates a general principle about interval exchange transformations and a conjecture about their cousins piecewise isometries: the growth of $a_V$ is sub-exponential (specifically, algebraic) because of the lack of stretching by the map.

Another way to define mixing is by using ideas from the ergodic theory of dynamical systems. There, (strong) mixing is succinctly defined (see, for example, Ref.~\onlinecite{sow06}, Sec.~3.7 or Ref.~\onlinecite{bfk06}) as
\begin{equation}
\lim_{n\to\infty} \Area\big(\mathbf{\Phi}^n(\mathcal{A}_1)\cap\mathcal{A}_2 \big) = \Area(\mathcal{A}_1)\Area(\mathcal{A}_2),
\label{eq:math_mixing}
\end{equation}
where $\mathcal{A}_{1}$ and $\mathcal{A}_{2}$ are any two subsets of $\mathcal{B}_0$ and $\cap$ denotes the {intersection} of two sets (that is, the material they have in common). Put simply, under iteration of the map $\mathbf{\Phi}$, the set $\mathcal{A}_1$ eventually becomes spread out evenly throughout the domain $\mathcal{B}_0$ so that no matter what other set $\mathcal{A}_2$ we choose, the amount of material in it that came from $\mathcal{A}_1$ is the same. In a more formal definition, $\Area(~)$ is replaced by the appropriate {invariant measure}. The maps we consider here are area-preserving, and therefore area is the proper measure of ``size'' for the definition.

Following the argument of Ref.~\onlinecite{sow06} (Theorem~3.7.2), suppose $\mathcal{A}_1$ is the gray region of the initial configuration $\mathcal{B}_0$ (recall Fig.~\ref{fig:maps}), that is, $\mathcal{A}_1 = [0,1]\times\left[0,\tfrac{1}{2}\right]$. Then, no matter what set we pick for $\mathcal{A}_2$, the number of stripes of each material within this volume grows exponentially, eventually resulting in half of each. Therefore, as $n\to\infty$, $\Area\big(\mathbf{\Phi}^n(\mathcal{A}_1)\cap\mathcal{A}_2 \big)\to\tfrac{1}{2}\Area(\mathcal{A}_2)$. Because $\Area(\mathcal{A}_1) = \tfrac{1}{2}$, it follows that the left- and right-hand sides in Eq.~\eqref{eq:math_mixing} are equal, and the baker's map represents strong mixing.

For the cutting and shuffling maps it should be clear that this argument breaks down. The second cutting and shuffling map (CS2) in Fig.~\ref{fig:maps} illustrates how the gray and white stripes under this map have non-uniform thickness and a pair of previously cut gray pieces may be ``glued back together'' at a later iteration. Thus, a cutting and shuffling map based on an interval exchange transformation can only be shown to be \emph{weakly mixing}\cite{af07}:
\begin{equation}
\lim_{n\to\infty} \frac{1}{n}\sum_{i=0}^{n-1} \big|\Area\big(\mathbf{\Phi}^i(\mathcal{A}_1)\cap\mathcal{A}_2 \big) - \Area(\mathcal{A}_1)\Area(\mathcal{A}_2) \big| = 0.
\label{eq:weak_math_mixing}
\end{equation}
This proof is far too involved to present here. Put simply, Eq.~\eqref{eq:weak_math_mixing} relaxes the requirement that the same amount of material from the subset $\mathcal{A}_1$ be found in any other subset $\mathcal{A}_2$. Instead, it is only required that this be true {on average, over many iterations}. Still, weak mixing is a stronger result than the ergodicity property mentioned previously that an interval exchange transformation acquires by satisfying the Keane condition.\cite{Note10}

To summarize, according to both the practical measure of mixing defined in Eq.~\eqref{eq:av} and the mathematical measures of mixing defined in Eqs.~\eqref{eq:math_mixing} and \eqref{eq:weak_math_mixing}, both of these protocols mix, albeit at different rates and with different ``strengths.'' In this respect, the cutting and shuffling map defies conventional wisdom -- there is no stretching, all Lyapunov exponents are zero, and the trajectories of material points in the continuum under this map are {not} chaotic.\cite{Note11}

\section{Inferring Stretching from Discrete Data\label{sec:numerics}}

\begin{figure}[!ht]
\includegraphics[width=0.75\columnwidth]{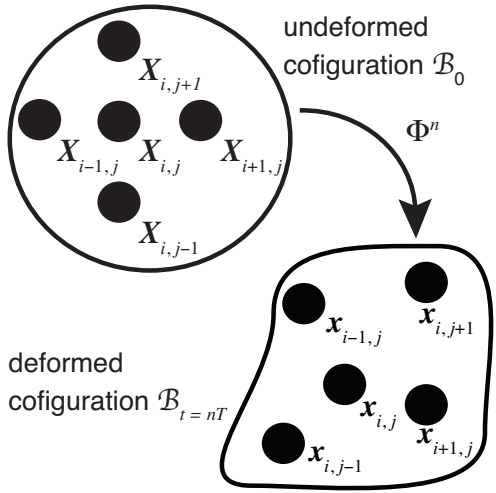}
\caption{Cumulative movement of material points of the undeformed continuum $\mathcal{B}_0$ after $n$ iterations of $\mathbf{\Phi}$, resulting in the deformed configuration $\mathcal{B}_{t=nT}$.}
\label{fig:numerical_dg}
\end{figure}

Suppose we performed an experiment with white and gray putty and recorded the initial and final configurations. We would like to find out how the putty was stretched due to the mixing protocols depicted in Fig.~\ref{fig:maps}. In a typical experiment only a finite number of material points in a continuum can be tracked. In other words, some collection of $M^2$ material points $\{\bm{X}_{i,j}\}_{i,j=1\hdots M}$ in $\mathcal{B}_0$ in the undeformed configuration are identified, and their locations $\{\bm{x}_{i,j}\}_{i,j=1\hdots M}$ in $\mathcal{B}_{t=nT}$ in various deformed configurations (for example, for different $n$) recorded, as shown schematically in Fig.~\ref{fig:numerical_dg}. From these values the motion and deformation of the continuum can be reconstructed.

A particularly simple and effective numerical approach to achieving this reconstruction is the standard central difference approximation to the deformation gradient [recall Eq.~\eqref{eq:dg_matrix}]:
\begin{equation}
\widetilde{\mathbf{F}}(\bm{X}_{i,j};n) = \begin{pmatrix} \dfrac{x_{i+1,j} - x_{i-1,j}}{X_{i+1,j} - X_{i-1,j}} & \dfrac{x_{i,j+1} - x_{i,j-1}}{Y_{i,j+1} - Y_{i,j-1}} \\ \dfrac{y_{i+1,j} - y_{i-1,j}}{X_{i+1,j} - X_{i-1,j}} & \dfrac{y_{i,j+1} - y_{i,j-1}}{Y_{i,j+1} - Y_{i,j-1}} \end{pmatrix} \approx \left(\nabla \mathbf{\Phi}^n\right)^\top\Big|_{\bm{X}=\bm{X}_{i,j}}.
\label{eq:central_dg}
\end{equation}
This approximation is used, for example, in extracting \emph{Lagrangian coherent structures} (barriers to mixing and transport) from experimental and simulated data,\cite{slm05_url} in determining the type (hyperbolic, elliptic, or parabolic) of periodic points (that is, $\bm{X}$ such that $\mathbf{\Phi}^p(\bm{X})=\bm{X}$, $p=1,2,\hdots$) of mixing protocols,\cite{kmpm99,mclo06} and in calculating the largest principal stretch in fluid mixing experiments with laminar\cite{vhb02,avg05} or turbulent\cite{tarvge08,div10} velocity fields. Understanding stretch fields is of immense practical importance in both the industrial and laboratory setting from the micro to the planetary scale. Their use in finding barriers to transport and mixing in the ocean was discussed in Ref.~\onlinecite{LCS_NYT}.

Tracking the motion of a collection of material points, as illustrated schematically in Fig.~\ref{fig:numerical_dg}, is the objective of this numerical approach. It may seem that many rearrangements of the continuum can result in a non-trivial deformation gradient, at least somewhere in the domain. Therefore, it would {appear} that stretching occurs under most maps. However, such intuition can often be wrong.
\begin{figure*}[!ht]
\includegraphics[width=0.75\textwidth]{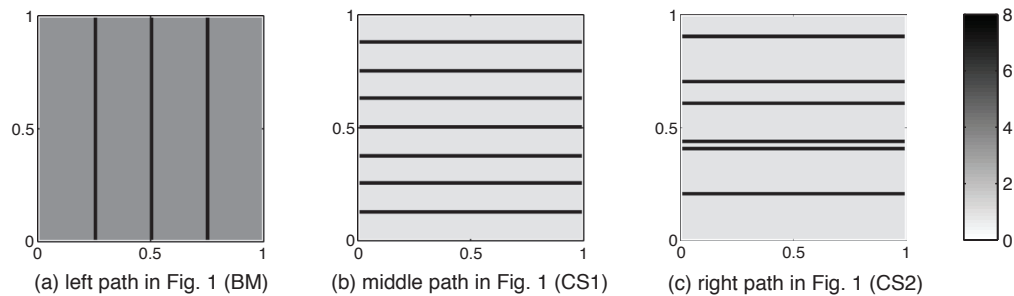}
\caption{Largest numerically-computed principal stretch field $\hat{\sigma}(\bm{X};n)$ for the protocols depicted in Fig.~\ref{fig:maps}. The background value is $4$ in (a) but equals $1$ in (b) and (c).}
\label{fig:stretches}
\end{figure*}

Figure~\ref{fig:stretches} shows the numerical results for $\hat{\sigma}$ based on tracking the movement of $M^2 = 251^2$ uniformly-distributed material points under each of the maps in Fig.~\ref{fig:maps} for the number of iterations of each map depicted there. Note that the three narrow vertical lines of high $\hat{\sigma}$ values for the baker's map and similar horizontal lines for the cutting and shuffling maps are due to the cuts in these protocols causing the stretch to be infinite there (that is, $\mathbf{F}^n$ is undefined). The lines are vertical for the baker's map, because the cuts are vertical. The middle line in Fig.~\ref{fig:stretches}(a) corresponds to the cut that is made in the first application of the map, and the other two lines correspond to the cut made in the second application of the map, which illustrates the set $\mathcal{C}_{\mathrm{BM}}^2$ defined in Eq.~\eqref{eq:baker_cuts}. For the cutting and shuffling maps, the lines are horizontal because the cuts are horizontal. Specifically, the dark horizontal lines in Fig.~\ref{fig:stretches}(b) depicts the set of cuts $\mathcal{C}_{\mathrm{CS1}}^1 = \{\bm{X}\;\text{in}\;[0,1]^2\,|\,0\le X\le1;\;Y = i/8,\; i=1,\hdots,7\}$, while those in Fig.~\ref{fig:stretches}(c) depict the set of cuts $\mathcal{C}_{\mathrm{CS2}}^n$ ($n=2$) noted after Eq.~\eqref{eq:iet_dgn}.

The apparent thickness of these lines is due to the coarse-grained view of the deformation that we obtain from the numerical data. Taking more points (larger $M$) makes these lines arbitrarily thin. (Their thickness is $\approx 2/M$, that is, twice the spacing between points because we used a central difference approximation to $\mathbf{F}$.) If we had applied the map more times (larger $n$) and/or had fewer material points, then the resulting thicker cuts would completely obscure the picture.

In Fig.~\ref{fig:stretches} the expected results are found. We applied the baker's map twice [recall Fig.~\ref{fig:maps} (BM)], so that $n=2$. From Eq.~\eqref{eq:baker_stretch}, the largest principal stretch is $\hat{\sigma} = 2^2 = 4$. We can also calculate $\hat{\sigma}$ from the numerical data by taking the square root of the largest eigenvalue of $[\widetilde{\mathbf{F}}_{\mathrm{BM}}(\bm{X};2)]^\top\widetilde{\mathbf{F}}_{\mathrm{BM}}(\bm{X};2)$ (recall the discussion at the end of Sec.~\ref{sec:kinematics}), where $\widetilde{\mathbf{F}}$ is defined in Eq.~\eqref{eq:central_dg}. In complete agreement with the theoretical result, the numerical calculation produces a ``background'' value (that is, the value of $\hat{\sigma}$ away from the cuts discussed above) of 4, as shown in Fig.~\ref{fig:stretches}(a). This result can also be anticipated by realizing that the baker's map stretches the continuum in the $X$-direction by a factor of 2. Applying the map twice gives a stretch ratio of 4. Similarly, for both of the cutting and shuffling maps, the largest principal stretch is equal to 1 for all $n$ from Eq.~\eqref{eq:iet_stretch}, that is, the maps do not stretch. The numerical calculation shown in Figs.~\ref{fig:stretches}(b) and (c) confirms that $\hat{\sigma}=1$. In both cases, the background value in the plots is 1, and $\hat{\sigma}$ is different from the ``background'' value only along the cuts, where it is infinite.

\section{Conclusion}

The mixing of continua can be accomplished by a great variety of maps of different complexities. In mechanics, the central theme is stretching. Stretching leads to a positive Lyapunov exponent and chaos when the problem is translated into the language of dynamical systems. For mixing, the central theme is generation of interfacial area per unit volume $a_V$. Stretching and folding (equivalently, chaotic dynamics) is an efficient way to do this.

Surprisingly, cutting and shuffling can also ``rearrange'' material points in a continuum quite well. The cutting and shuffling maps, which would be considered pathological in ``classical'' continuum mechanics,\cite{t91,g81} do not stretch, possess no positive Lyapunov exponents, and exhibit no chaotic behavior in the usual sense, yet they mix. Even though interfacial area is produced at an {asymptotically} slower rate than for a chaotic motion that stretches, for short times it appears possible for cutting and shuffling to dominate stretching and folding, depending on the number of cuts $k$ in Eq.~\eqref{eq:iet_av} and the growth rate $h$ in Eq.~\eqref{eq:bakers_av}. Although we may be misled to conclude that a cutting and shuffling map stretches the underlying continuum based on the complicated pattern of non-uniform striations produced by it, analytical and sufficiently-refined numerical calculations show otherwise.

Recent work\cite{smow08,col10,col10b,jlosw10} takes the idea of cutting and shuffling even further, arguing that it provides the ``skeleton'' of certain regimes of granular flow in tumblers. In the present work, we illustrated how cutting and shuffling can also lead to mixing without chaos. And, in more complicated maps, cutting and shuffling can lead to complex dynamics without chaos in any usual sense of the word.\cite{smow08,col10,col10b,jlosw10} Thus, cutting and shuffling maps and their generalizations are more than just a mathematical exercise, and have the potential to accurately describe the underlying framework of mixing in physical systems.

\begin{acknowledgments}
We would like to thank Howard Stone for bringing to our attention the work on card shuffling.\cite{ad86,tt00} Marissa Krotter provided valuable advice and numerical results on the mixing properties of interval exchange transformations. Incisive criticism from Stephen Wiggins and Rob Sturman improved the paper. The careful reading and helpful comments by the reviewers are also much appreciated. I.C.C.\ was supported by a Walter P.\ Murphy Fellowship from the Robert R.\ McCormick School of Engineering and Applied Science at Northwestern University.
\end{acknowledgments}

\end{document}